\documentclass[aps,pre,twocolumn, titlepage,showpacs]{revtex4}

\usepackage{graphicx}% Include figure files
\usepackage{color}
\usepackage{dcolumn}% Align table columns on decimal point
\usepackage{float}
\usepackage{bm}% bold math

\begin{document}

\title{Mobility in a strongly coupled dusty plasma with gas}

\author{Bin Liu}
\author{J. Goree}
\affiliation{Department of Physics and Astronomy, The University
of Iowa, Iowa City, Iowa 52242}
\date{\today}

\begin{abstract}

The mobility of a charged projectile in a strongly coupled dusty plasma is simulated. A net force $F$, opposed by a combination of collisional scattering and gas friction, causes projectiles to drift at a mobility-limited velocity $u_p$. The mobility $\mu_p=u_p/F$ of the projectile's motion is obtained. Two regimes depending on $F$ are identified. In the high force regime, $\mu_p \propto F^{0.23}$, and the scattering cross section $\sigma_s$ diminishes as $u_p^{-6/5}$. Results for $\sigma_s$ are compared with those for a weakly coupled plasma and for two-body collisions in a Yukawa potential. The simulation parameters are based on microgravity plasma experiments.

\end{abstract}

\pacs{47.55.D-, 47.60.-i, 47.20.Ky, 63.22.-m}\narrowtext

\maketitle

\section{Introduction}
\label{sec:intro}

A projectile driven by a net force $\textbf{F}$ through a medium of target particles will collide with them, and it will drift in the direction parallel to $\textbf{F}$ at an average velocity $u_{p}$. This motion is described by the transport coefficient for mobility
\begin{equation}\label{mobility}
    \mu_{p}=u_{p}/F.
\end{equation}

The target particles can be in any state of matter.  Research on mobility and diffusion of electrons and ions began over 100 years ago for {\it gases}~\cite{Thomson:1906}, and later for {\it solids}~\cite{Seitz:1948,Kittel:1976} and {\it weakly coupled plasmas} \cite{Spitzer:1953,Braginskii:1965}.

Here, the target we investigate is a {\it strongly coupled} plasma, in which the potential energy exceeds the kinetic energy, so that particles self-organize into a liquid-like or solid-like structure~\cite{Ichimaru:1982}. Strongly coupled plasmas in nature include neutron star crusts~\cite{Horowitz:2007}, giant planet interiors, and white dwarf interiors~\cite{Kalman:1998}. Strongly coupled plasma can be realized in the laboratory using a dusty plasma, which is a four-component mixture of electrons, ions, neutral gas, and  micron-size particles of solid matter~\cite{Thomas:1994,Chu:94,Melzer:96,Juan:1998,Shukla:02,Liu:03,Feng:2007,Ishihara:2007,Morfill:09,Melzer:08,Flanagan:2009,Fortov:2009,Piel:2010,Bonitz:10}. The solid particles, which we call dust particles, become strongly coupled due to their large charges.

We investigate a system that hinders a projectile's motion by two types of collisions: Coulomb collisions among strongly coupled dust particles, and the friction due to collisions of gas atoms with the projectile. The latter is modeled as a simple drag term, which does not require a particle description of the gas atoms. The gas friction has been reviewed in~\cite{Epstein:1924, Liu:03}, and binary Coulomb collisions for an {\it isolated pair} of dust particles are reviewed in~\cite{Khrapak:2004}. In a strongly coupled plasma, Coulomb collisions are different from collisions of an isolated pair (i.e., binary) because the target particle in a strongly coupled plasma does not move freely as it recoils. Instead it collides immediately with other target particles, which collide with others in a chain of collisions. In this way, the Coulomb collisional process is collective and not binary~\cite{Baalrud:2013}. To simulate this system, we require a model that represents dust particles as discrete particles.

Since the collisions are so different in weakly and strongly coupled plasmas, one would expect transport coefficients, such as mobility, to be different as well. The velocity relaxation rate, which is related to the mobility, has been studied in ultracold plasmas with an ionic Coulomb coupling parameter $\Gamma$ of order unity~\cite{Bannasch:2012,Killian:2007}. The mobility and drift motion have also been studied in several two-dimensional strongly coupled Coulomb systems, which are not plasmas but have similar Coulomb collisions; these include colloidal crystals~\cite{Reichhardt:2004}, and electrons~\cite{Glasson2001,Kono2002,Ikegami2012} and ions~\cite{Barengi1991} on the surface of liquid helium.  To the best of our knowledge, mobility has not been studied much in strongly coupled plasmas with liquid-like conditions $\Gamma>10$, three-dimensional Yukawa systems, or dusty plasmas. Other transport processes including diffusion~\cite{Vaulina:02,Ratynskaia:2010,Dzhumagulova:12},  viscosity~\cite{Murillo:2001,Fortov:2012,Donko:06,Donko:08}, and thermal conductivity~\cite{Donko:04} have been studied for dusty plasmas. We expect mobility in a dusty plasma to be determined by two effects experienced by the dust particles: Coulomb collisions (which in dusty plasmas are modeled by a Yukawa potential) and frictional drag on the ambient neutral gas. The conditions we investigate are at a moderate value of $\Gamma$ where the strongly coupled plasma is in a dense liquid-like state.

There are at least two regimes of projectile transport, depending on the driving force $F$. In what we term the {\it low regime}, $F$ is small so that projectiles are near thermal equilibrium with target particles. In what we term the {\it high regime}, $F$ is so large as to cause a considerable departure from the thermal equilibrium.

The literature for ions in gases is well developed, and many experiments have been reported~\cite{Mason&McDanile,Viehland:1995,Wannier:1952}. It is known for that system that the transport in the high regime is different from the low regime: the mobility is not constant but varies with $F$ in a way that depends on the scattering potential~\cite{Dutton:1975,Viehland:1995}.

For the denser physical system of liquids instead of gases, while it is possible to propel a small projectile, the target's high density poses a great difficulty for attaining a superthermal speed for the projectile. Consequently, it is difficult to perform experiments to study mobility in a liquid in a high regime. This difficulty can be overcome by using a dusty plasma as a model system for a liquid because a dusty plasma has a small volume fraction~\cite{Feng:12}.

Motion of projectile dust particles through a cloud of target dust particles has been observed in recent microgravity dusty plasma experiments~\cite{Sutterlin:2009,Sutterlin:2010,Fink:2011,Caliebe:2011,Arp:2011,Schwabe:2011,Zhukhovitskii:2012}. For these observations, the target and projectile particles generally have different sizes. Here we simulate drifting motion as in the experiments of~\cite{Sutterlin:2009,Sutterlin:2010,Fink:2011}, except that we consider individual projectiles, not dense beams of projectiles, in order to determine a projectile's mobility coefficient due to collisions, without any cooperative motion among projectiles. A projectile drifts through a target due to a net force $F$; this net force could be due to an imbalance of electric and ion drag forces, as can happen for different dust particle sizes. Due to their different sizes, a projectile particle drifts, while the target particles are in a force equilibrium and do not drift. This situation is possible because of different scalings of forces with a particle's size~\cite{Samsonov:1999}.

In this paper, our main results are: (1) a characterization of two regimes of projectile transport, (2) an evaluation of mobility coefficient $\mu_p$ for projectiles, and (3) a determination of the scattering cross section $\sigma_s$ as a function of the drift velocity $u_p$.

\section{Simulation}
\label{sec:simulation}

We perform a three-dimensional (3D) Langevin molecular dynamics simulation of dust particle motion including Coulomb collisions. Dust particles also experience frictional drag on the gas atoms. Due to their charge $Q$, dust particles also repel one another with a Yukawa potential, $\phi(r)=Q^2\text{e}^{-r/\lambda_{D}}/4\pi\epsilon_0r$, where the screening length $\lambda_D$ due to electrons and ions reduces the interaction at a large distance of $r$.  This many-particle Yukawa system is described by dimensionless parameters
\begin{equation}\label{gamma}
    \Gamma ={{Q}^{2}}/4\pi{{\epsilon}_{0}}a{{k}_{B}}{{T_t}},
\end{equation}
where $k_BT_t$ is the kinetic temperature of the system,  and
\begin{equation}\label{kappa}
    \kappa=a/\lambda_{D} ,
\end{equation}
where
\begin{equation}\label{wigner}
    a =(3/4\pi n_t)^{1/3}
\end{equation}
is the Wigner-Seitz radius, and $n_t$ is the number density of dust particles. For microgravity experiments, typical parameters are $n_t=5\times{{10}^{4}}~\text{c}{{\text{m}}^{-3}}$~\cite{Arp:2010} and $a = 0.017$~cm.

We integrate the equations of motion~\cite{Ratynskaia:2009,Klumov:2009}
\begin{eqnarray}\label{EOM}
  m_{t}\ddot{\mathbf{x}_{i}} &=& -\nu_t m_{t}\dot{\mathbf{x}_{i}}+\gamma\mathbf{\zeta}_{ti}(t)-{\scriptstyle\sum_{k}\nabla}\phi_{ik}-{\scriptstyle\nabla}\Phi  \\
  m_{p}\ddot{\mathbf{x}_{j}} &=& -\nu_p m_{p}\dot{\mathbf{x}_{j}}+\gamma\mathbf{\zeta}_{pj}(t)-{\scriptstyle\sum_{k}\nabla}\phi_{jk}-{\scriptstyle\nabla}\Phi+\textbf{F}
\end{eqnarray}
for target and projectile particles, respectively. A constant net force $\textbf{F}=F\hat{x}$ acts only on the projectile. The first two terms on the right hand side are the frictional force with a coefficient $\nu$ and the Markovian fluctuating force $\zeta(t)$; both of these are due to collisions of gas atoms of temperature $T_{\rm gas}$ with dust particles. The fluctuating force has an amplitude set by the fluctuation-dissipation theorem, $\left\langle\zeta(t)\zeta(0)\right\rangle=2\nu mk_{B}T_{\rm gas}\delta(t)$. We integrate the equations of motion using an algorithm that incorporates the friction and the fluctuating force~\cite{Gunstern:1982}.
To account for particle heating mechanisms in addition to gas-atom collisions, we augment the Markovian fluctuating force by a multiplier $\gamma$~\cite{Goree:2013,note:3}, which would be unity for thermal equilibrium.
The terms in Eqs.~(\ref{EOM}) and (6) with gradients are the electric force due to particle-particle interaction $-{\scriptstyle\nabla}\phi$ and confinement $-{\scriptstyle\nabla}\Phi$. To simulate a 3D dusty plasma with a uniform spatial distribution, we choose a confining potential $\Phi$ that is mostly flat, with a rising parabola at the edge. Projectiles  introduced at the edge are spaced sufficiently so that they interact only with target particles and not with other projectiles, as demonstrated in Appendix~\ref{sec:simulationMethod}.

The net force $F$ can arise physically from an imbalance of the ion drag force and other forces, because the ion drag force depends on particle size~\cite{Khrapak:2002}. In this paper, we treat $F$ simply as an adjustable input parameter, which we vary over a wide range bracketing the values we expect in an experiment.

We use simulation boxes of two sizes. A larger force $F$ requires the larger box since the projectiles move a greater distance. We verified the simulation generates the same results with both box sizes in the range $6.8<F<10$. The box dimensions are: $132\times81\times69.3~\lambda_D^3$ for the smaller and $263\times122\times104~\lambda_D^3$ for the larger boxes. Boundary effects, such as the initial acceleration of the projectile when it is released, are avoided by analyzing data only in the central volume that excludes the edges.  Further details of the simulation method are in Appendix~\ref{sec:simulationMethod}.

Our simulation parameters are motivated by ground-based~\cite{Khrapak:2005} and microgravity~\cite{Fortov:2005, Fink:2011} experiments with the PK-4 instrument. The polymer particles have a density 1.51~g/cm$^{3}$. The projectiles have a radius $0.64~\mu\text{m}$ while the targets are $3.43~\mu\text{m}$ radius with mass $m_t=2.55\times {{10}^{-13}}{\text{kg}}$.  For neon at 50 Pa pressure, the ion and gas temperatures are assumed to be $0.03~\text{eV}$, and the electron density and temperature are estimated as $2.4\times {{10}^{8}}~{{\text{cm}}^{-3}}$ and $7.3~\text{eV}$~\cite{Goree:2013}, so that ${{\lambda}_{D}}={{(\lambda_{De}^{-2}+\lambda_{Di}^{-2})}^{-1/2}}=8.3\times {{10}^{-3}}~\text{cm}$.
%, where ${{\lambda}_{De}}$ and ${{\lambda}_{Di}}$ are the electron and %ion Debye lengths.
Our projectile particle charge is $Q_p =-1590~e$, based on Fig.~7(a) in~\cite{Khrapak:2005}, and our target particle charge is $Q_t=-8520~e$. The gas friction coefficients~\cite{Liu:03, footnote:2} are $\nu_p =273~{{\text{s}}^{-1}}$ and $\nu_t =51~{{\text{s}}^{-1}}$. The characteristic time for collective motion in the target is $\omega_t^{-1}$, where
\begin{equation}\label{omega}
    \omega _{t}=\sqrt{Q_t^{2}n_{t}/\epsilon_{0}m_{t}},
\end{equation}
which has a value of $157~\text{s}^{-1}$.

\section{Target conditions}

Since transport can vary with temperature, we perform simulations for two target temperatures, $T_t = 10T_m$ and $2T_m$, corresponding to $\Gamma = 62$ and $310$, respectively. Here, $T_m$ is the melting point~\cite{Hamaguchi:1997}. These two kinetic temperatures, which are $T_t=8.3$ and 1.66~eV in physical units, are achieved by selecting the multiplier $\gamma=16$ and 7, respectively. For all our simulations, $\kappa = 2.4$, corresponding to $n_t=3\times{{10}^{4}}~\text{c}{{\text{m}}^{-3}}$ and $a=0.02$~cm.

\begin{figure}
\centering
\includegraphics{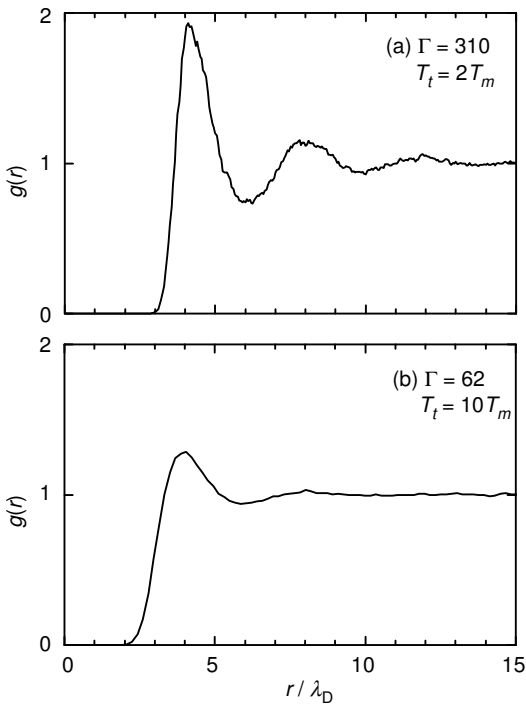}
\caption{\label{Gr} Characterization of simulation conditions for $\kappa = 2.4$ at two temperatures: (a) $\Gamma = 310$ or $T_t=2T_{m}$ and (b) $\Gamma=62$ or $T_t=10T_{m}$. The pair correlation functions shown here indicate that the target has a liquid-like structure.}
\end{figure}

To characterize the target, we performed a simulation without projectiles. Figure~\ref{Gr} shows the pair correlation function $g(r)$ from our simulation for these two conditions.

The 3D structure of the target, for $T_t=2T_m$, can also be viewed from a movie which we include in the Supplemental Material~\cite{EPAPS}. This movie shows a still image of the three-dimensional structure, viewed from a time-varying angle.

As the projectile moves through the target there is a shear motion on a microscopic scale, i.e., a scale analogous to the molecular scale in a simple liquid. If the shear motion were instead on a macroscopic or hydrodynamic scale, with a gradient length of at least a dozen interparticle spacing~\cite{Tabeling:2005}, the target's collective behavior could be described by its viscosity. We determined this viscosity, using the standard Green-Kubo method~\cite{Hansen&McDonal:1986}, to have a value 0.065 and $0.044~n_{t}{{m}_{t}}{{a}^{2}}{{\omega }_{t}}$ for $T_t = 2T_m$ and $10 T_m$, respectively. In physical units, these viscosities are $3.1\times10^{-9}$ and $2.1\times10^{-9}$~g~mm$^{-1}$~s$^{-1}$. Later we will make use of the idea that the viscosity is lower at higher temperatures.

\begin{figure}[htb]
\centering
\includegraphics{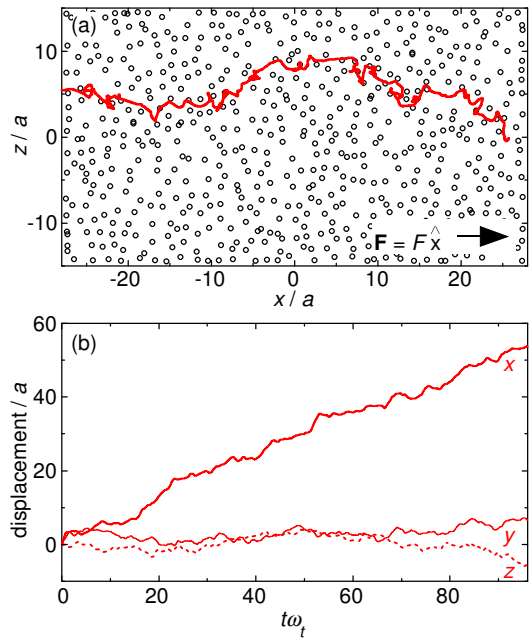}
\caption{\label{orbit} (Color online) (a) A typical projectile trajectory shown as a curve projected onto the $x-z$ plane, from a run at $T_t=10T_m$. Also shown is a snapshot of target particle positions within a slab of thickness $\Delta y = 1.7a$. (b) Time series of displacements of a representative projectile, showing drift in the $\hat{x}$ direction and random walk or diffusion in the $\hat{y}$ and $\hat{z}$ directions. Data shown are for $F = 3.8$. The time series duration corresponds to 610~ms in physical units.}
\end{figure}

\section{Results}
\label{sec:results}

We present our results in dimensionless units. We normalize distance, time, velocity, force, temperature, and mobility by $a$, $\omega_t^{-1}$, $a\omega_{t}$, $m_p\omega_t^2a$, $m_t(\omega_{t}a)^2$, and $(m_p\omega_{t})^{-1}$, respectively.

\begin{figure}[htb]
\centering
\includegraphics{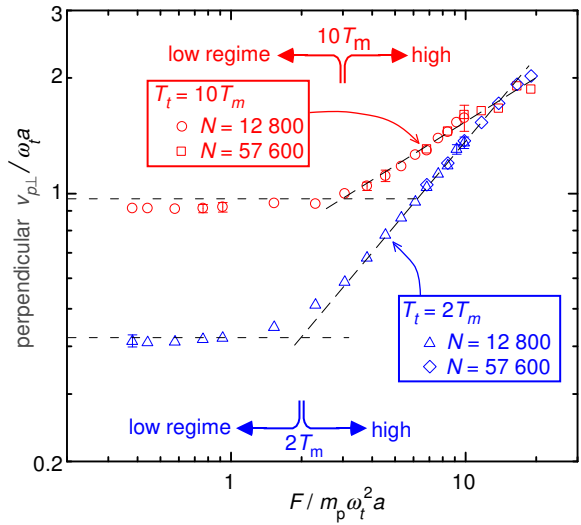}
\caption{\label{velocity} (Color online) Characterization of regimes using projectile's random speed $v_{p\perp}$ in the direction $\perp \textbf{F}$. Two regimes are seen and the transition between them is identified by the intersections of the asymptotes (dashed lines). Speed is normalized here by $\omega_t a$, which has a value 31.4~mm/s. Simulations were performed with two sizes $N$ for the number of target particles.}
\end{figure}

The projectile motion, Fig.~\ref{orbit}(a), reveals the drift parallel to $\textbf{F}=F\hat{x}$, and random scattering in the perpendicular direction. In Fig.~\ref{orbit}(b), the projectile's drift is seen in the time series for the displacement $x$, which has a slope that corresponds to the drift velocity. The perpendicular displacements $y$ and $z$ exhibit only a random walk.

We calculate the perpendicular random velocity $v_{p\perp}=(\dot{x}^2+\dot{y}^2)^{1/2}$, and we calculate the parallel drift velocity $u_p$ by fitting the $x$ displacement as in Fig.~\ref{orbit}(b) to a straight line. Results for $v_{p\perp}$ and $u_p$ are presented in Fig.~\ref{velocity} and Fig.~4(a), respectively. These velocity results are presented using log-log axes so that we can identify power-law scalings. We will next use the magnitude of $v_{p\perp}$ to identify regimes of the projectile motion, and after that we will use the drift velocity $u_p$ to determine the mobility $\mu_p$ and the scattering cross section $\sigma_s$.

\subsection{Characterization of regimes}
\label{subsec:regimes}

As our first chief result, we will identify the transition between regimes of the projectile's motion. In the high regime, the perpendicular random velocity $v_{p\perp}$ increases with $F$, as projectiles gain significant random energy from the acceleration corresponding to $F$, while in the low regime $v_{p\perp}$ has a constant value, Fig.~\ref{velocity}.

We identify the transition between regimes as the intersection of asymptotes in Fig.~\ref{velocity}. The force at the transition is found to be $F \approx 2$ or 3, as marked with arrows in Fig.~\ref{velocity}, for $T_t=2T_m$ or $10T_m$, respectively. We note that these values for the transition coincide with the conditions that yield a drift velocity comparable to the equilibrium thermal velocity of the projectile, $u_p\approx\sqrt{k_BT_t/m_p}$. The latter finding is comparable to the case for ion projectiles in a gas~\cite{Wannier:1952}.

\begin{figure}[h!]
\centering
\includegraphics{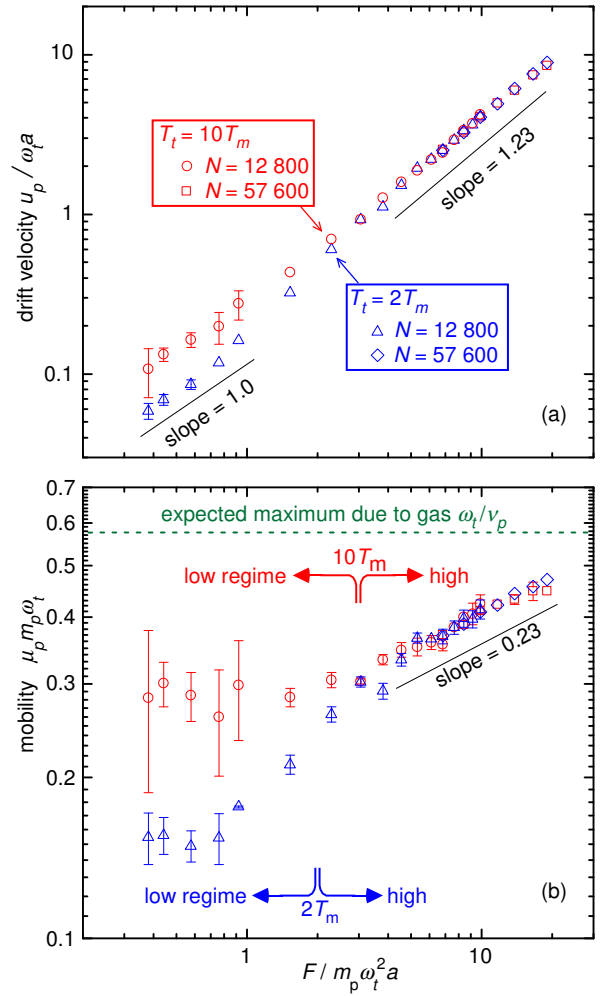}
\caption{\label{drift} (Color online) (a) Projectile speed $u_p$ in the direction $\parallel \textbf{F}$. This drift velocity scales as $u_p\propto F^{1.01\pm0.12}$ in the low-force regime  and $u_p\propto F^{1.23\pm0.02}$ in the high-force regime. (b) Mobility dependence with $F$. In the high regime (large $F$), we find  $\mu_p \propto F^{0.23}$. We expect the maximum mobility limit to be $(m_p\nu_p)^{-1}$, as indicated by the dashed line, corresponding to the gas drag on a projectile without Coulomb collisions. The power law scaling of the mobility is the same for two temperatures we simulated.}
\end{figure}

\subsection{Evaluation of mobility coefficient}
\label{subsec:coefficients}

To determine the mobility $\mu_p=u_p/F$, which is our second chief result, we divide the drift velocity $u_p$ in Fig.~\ref{drift}(a) by the force $F$, which is the horizontal axis in that graph. The resulting mobility data are presented in Fig.~\ref{drift}(b). The mobility typically has a value in the range 0.16 to $0.5~(m_p\omega_t)^{-1}$, for the target temperatures and range of forces that we consider. In physical units, this range corresponds to $6.1\times10^{8}$ to $1.92\times10^{9}$ ${\rm g}^{-1}{\rm s}$ for the PK-4 parameters listed in Sec.~\ref{sec:simulation}. If there were no Coulomb collisions to retard the motion of the drifting projectile, the mobility would be limited only by gas friction and it would have a limiting value of $0.58~(m_p\omega_t)^{-1}$, as indicated by the dashed line. All our data points from the simulation lie below this limiting value due to the combination of Coulomb collisions and gas friction, which both retard the projectile's motion in response to the force $F$.

A power-law scaling for the mobility can be found by noting that data lie mostly on straight lines, in the log-log plots of Fig.~\ref{drift}.  By fitting, we find that $u_p$ varies as $\propto F^{1.23\pm0.02}$ in the high regime, where nonequilibrium effects become significant, as compared to the scaling $F^{1.01\pm0.12}$ for the low regime. Correspondingly, the mobility $u_p/F$ is essentially constant in the low regime, while it has an exponent  of 0.23, i.e., $\mu_p \propto F^{0.23}$, in the high regime. Expressing the scaling in terms of drift velocity instead of force, we find $\mu_p \propto u_p^{0.19}$ in the high regime.

We expect that these scaling laws for the mobility will fail at even higher forces because the mobility cannot exceed the limiting value due to gas friction. This limiting value is $(m_p\nu_p)^{-1}$, which is $0.58(m_p\omega_t)^{-1}$ for a particle of 0.64~$\mu {\rm m}$ radius in a 50 Pa Neon gas. This limit is, in effect, a third regime, which we did not explore because it would require forces that we expect to be unattainably large in experiments such as PK-4. However, we expect an analogous limit must occur in a colloid due to friction on the solvent, and that limit might be easily attained because of the stronger friction effect for a liquid solvent, as compared to the rarefied gas in a dusty plasma.

The target temperature is found not to have an effect on the mobility in the high regime. This result is seen by the overlapping data points in the right hand side of Fig.~\ref{drift}(b), where the mobility obeys the same $\mu_p \propto F^{0.23}$ power law for both temperatures.

Temperature does, however, affect the constant value of the transport coefficients in the low regime. This is seen on the left side of Fig.~\ref{drift}(b), where we find $\mu_p=0.16\pm 0.01$ for $T_t = 2T_m$, which is different from $\mu_p=0.29\pm0.02$  for $T_t = 10T_m$.

We can speculate why, in the low regime, $\mu_p$ is lower for our colder temperature. As mentioned earlier, the disturbance created amongst the target particles by the moving projectile is like a shear motion with a microscopic scale. If it instead had a macroscopic scale, the shear motion could be described by a hydrodynamic equation where shear motion is opposed by dissipation characterized by a shear viscosity. It is well known~\cite{Hamaguchi:2002} that for a strongly coupled plasma the shear viscosity varies oppositely with $T_t$ when $T_t$ is only a modest multiple of $T_m$ as it is in our case. Even though we can not apply the hydrodynamic equations to the microscopic shear in our target, we expect the same tendency of the shear motion to experience a greater dissipative resistance at a colder temperature. This expected tendency agrees with our finding that $\mu_p$ increases with $T_t$.

\begin{figure}[h!]
\centering
\includegraphics{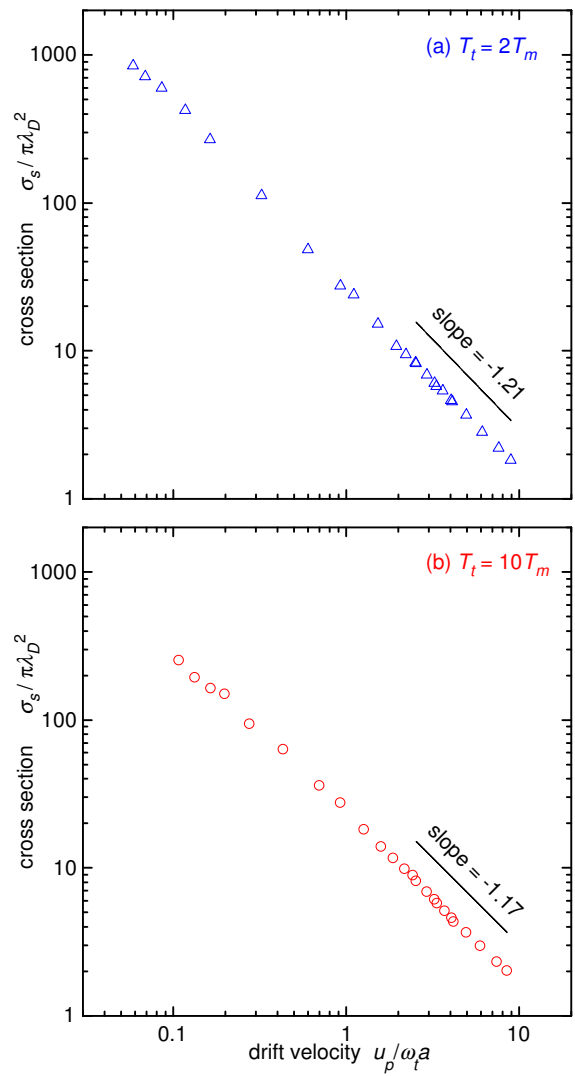}
\caption{\label{scaling} (Color online) Scattering cross section $\sigma_s$ for target at different temperatures (a) $T_t = 2T_m$ and (b) $T_t=10T_m$. The scattering cross section is calculated from Eq.~(\ref{nu_pt}) using the results in Fig.~4(b) for the mobility, which includes the effects due to gas friction. The cross section exhibits a power law scaling, which approaches $\sigma_s\propto u_p^{-6/5}$ at large drift velocity, for both temperatures we simulated.}
\end{figure}

\subsection{Determination of the scaling of $\sigma_s$}
\label{subsec:scaling}

As our third chief result, we find the slowing-down cross section $\sigma_s$, which is also often called a momentum transfer cross section~\cite{Mason&McDanile}. We use the force balance equation $\nu_{pt}m_p u_p = F = u_p / \mu_p$ for a projectile moving at a constant drift velocity $u_p$, where $\nu_{pt} = n_t \sigma_s u_p$ is the collision frequency for projectiles to slow down. Combining these equations with Eq.~(\ref{wigner}) yields an expression for $\sigma_s$
\begin{equation}\label{nu_pt}
    \sigma_s=\frac{4\pi a^2}{3}\left(\frac{a\omega_t}{u_p}\right)\frac{1}{m_p\omega_t\mu_p},
\end{equation}
which we will use to obtain $\sigma_s$ from our results for $u_p$ and $\mu_p$.

Results for $\sigma_s$ are presented in Fig.~\ref{scaling} as a function of the drift velocity $u_p$. The cross section diminishes with $u_p$, and in the log-log plots the data fall mostly on a straight line, indicating that $\sigma_s$ obeys a power law. The power law scalings, obtained by fitting the data in the high regime, are $\sigma_s\propto u_p^{-1.21\pm0.02}$ for $T_t=2T_m$ and $\sigma_s\propto u_p^{-1.17\pm0.02}$ for $T_t=10T_m$. The exponent in both cases is $\approx-6/5$. We will next compare this exponent for our many-body collective system to the exponent for two binary systems.

For the familiar binary system of a fast projectile scattering in a $1/r$ Coulomb potential, which is the case for a weakly coupled plasma, the exponent is  $-4$, i.e., $\sigma_s\propto u_p^{-4}$. Our exponent of $-6/5$ is a much weaker dependence. The system we simulate is different in three ways. Instead of the binary small-angle collisions that are typical of a weakly coupled plasma,  we have large angle scattering and collective effects among the target particles, which collide with one another as they recoil. Our scattering potential is Yukawa instead of $1/r$. Finally, our system includes dynamical friction with gas atoms.

\begin{figure}[h!]
\centering
\includegraphics{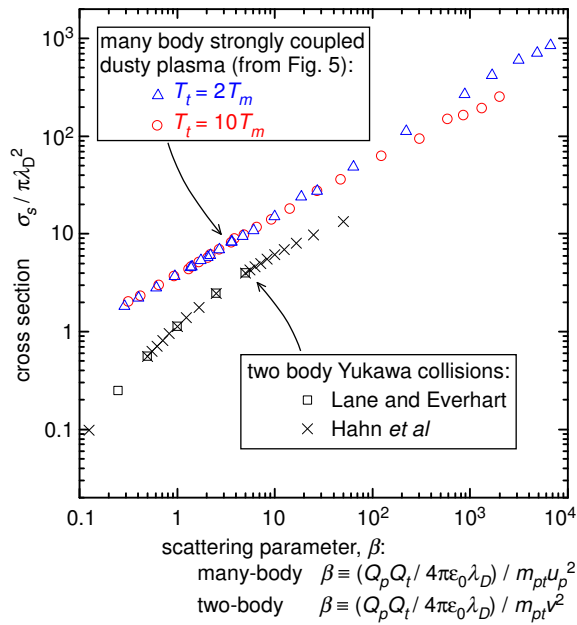}
\caption{\label{comparison} (Color online) Comparison of the scattering cross section for a strongly coupled dusty plasma (our data for $T_t=2T_m$ and $10T_m$) with that for classical two-body collision in a Yukawa potential by Lane and Everhart~\cite{Lane:1959} and Hahn {\it et al}~\cite{Hahn:1971}. In the range of $0.2<\beta<50$, the cross section in our strongly coupled many-body dusty plasma is generally larger than that for the two-body collision; it also exhibits a single power-law scaling with $\beta$.}
\end{figure}

Another binary system for comparison is a projectile that is scattered by an isolated target which has a Yukawa potential. This was also studied long ago~\cite{Lane:1959,Hahn:1971}, without gas. In Fig.~\ref{comparison}, we replot our cross-section data to compare with the binary-Yukawa data from Table II of Ref.~\cite{Lane:1959} and Table I of Ref.~\cite{Hahn:1971}. As in Ref.~\cite{Khrapak:2004}, we normalize the cross section by $\pi\lambda_D^2$, and the horizontal axis represents the scattering parameter,
\begin{equation}\label{beta}
    \beta(v)=\frac{Q_pQ_t}{4\pi\epsilon_0\lambda_D}\frac{1}{m_{pt} v^2}~,
\end{equation}
where $m_{pt}=m_pm_t/(m_p+m_t)$ is the reduced mass, and $v$ is the relative velocity before collisions. For our data, we replace the relative velocity $v$ (for the binary system) with the drift velocity $u_p$ (which is suitable for the many-body target).

Based on the comparison in Fig.~\ref{comparison}, we find that the scattering cross section for our strongly coupled dusty plasma differs from that of classical two-body collisions in a Yukawa potential in two ways. First, the cross section for our dusty plasma is generally larger than that of the two-body collision. Second, our data tend to exhibit a distinct power-law scaling for $\sigma_s$ vs $\beta$, unlike the two-body case, where $\sigma_s$ does not follow a single power law scaling with $\beta$. These differences can arise from two effects that are present in the dusty plasma but not the binary Yukawa case: gas friction and collective effects in the collisions in a strongly coupled plasma system, in which the motion of a recoiling particle is hindered by interactions with neighboring target particles.

\section{Summary}
\label{sec:summary}

In summary, we investigated a charged projectile drifting through a dusty plasma, taking into account two processes that are significant in experiments: Coulomb collisions in a many-body strongly coupled dusty plasma, and gas friction. We determined the mobility for the projectile and characterized the two regimes of projectile motion. For this strongly coupled plasma, the scaling of $\mu_p$ with $F$ in the high regime indicates a scattering cross section $\sigma_s\propto u_p^{-6/5}$ in the range of force we studied. Our results for $\sigma_s$ are larger than that for two-body collisions in a Yukawa potential in the absence of gas. We anticipate that mobility-limited drift of an isolated projectile through a target of strongly coupled dusty plasma can be observed in future dusty plasma experiments using video imaging. The experiment would require that the projectile has a different size from the target, so that there is a net force that can drive the projectile while the target particles remain in a non-drifting equilibrium.

Remaining issues that could be addressed in future work include the dependence of projectile motion on target parameters such as $\Gamma$ and $\kappa$, the relationship between various transport coefficients, and the possibility of extending our work to other systems such as a Yukawa one component plasma (YOCP)~\cite{Ohta:2000, Daligault:2012,Rosenfeld:1999}.

\appendix
\section{Simulation method}
\label{sec:simulationMethod}

Here we provide further details of the simulation method.

\subsection{Confinement}

We model a small portion of a 3D dusty plasma by confining particles in a finite rectangular volume. The confining potential is flat in most of the volume, and a rising parabola at the edge, i.e.,
\begin{equation}\label{confinement}
\Phi=\psi(x,b)+\psi(y,c)+\psi(z,d),
\end{equation}
where
\begin{equation}\label{confiningPotential}
\psi(x,b) =
\cases{
  0, & $|x|< b$ \cr
  m_t\omega_e^2(|x|-b)^2/2, & $|x|\ge b$ \cr
}
\end{equation}
and similarly for $y$ and $z$. The main volume, where we analyze our results, has a flat potential, $\psi =0$,
with a width $2b$, $2c$, and $2d$ along
the $x$, $y$, and $z$ axes, respectively. Here, $\omega_{e}$ is a
constant that characterizes the parabolic confinement at the edge. The design of
this confining potential helps provide a number density that is
uniform everywhere except within $7{{\lambda }_{D}}$ of the edge,
according to our simulation test, with the constant $\omega_{e}$
chosen to be $\sqrt{Q_t^2/4\pi\epsilon_{0}m_{t}\lambda_{D}^3}$. To avoid any boundary
effects, in our analysis we will use data only from the central
portion of the simulated volume, i.e., $|x_{i}|\le 0.84b$,
$|y_{i}| \le 0.86c$, and $|z_{i}|\le 0.86d$. We perform our simulation with two system sizes, $N=12~800$ and $57~600$ target particles, and we found no significant size effect.

\subsection{Potential truncation}

For efficiency, we truncate the Yukawa potential at a large cutoff radius of $13.25\lambda_{D}$. At this distance the potential is five orders of magnitude smaller than at the distance of a nearest neighbor.

\subsection{Initial configuration}

We perform four simulation runs for each value of the force $F$. Each run is done with a different initial configuration of the target particles. For each initial configuration, we record time series of particle positions and velocities for a duration of $480~\omega _{t}^{-1}$.

\subsection{Integration}

We numerically integrate the equations of motion, Eqs.~(\ref{EOM}) and (6), using the Langevin integrator of~\cite{Gunstern:1982}. To account for disparate time scales for the lighter projectile and heavier target particles, we use a multiple-time-scale method~\cite{Tuckerman:1991}.

Our time steps, $2.3\times10^{-4}~\omega_{t}^{-1}$ and $4.5\times10^{-6}~\omega_{t}^{-1}$ for the target and projectile particles, respectively, were selected by performing a convergence test. In the convergence test, we solved $m\ddot{x_{i}}=-\nabla\phi_{ij}-\nabla\psi$ for a system consisting of only two particles. A projectile was directed toward a stationary target particle with zero impact parameter. Because of the confinement $\psi$, these particles repeatedly collided. We calculated the discrepancy in a particle's position and varied the time step downward until the discrepancy was $<0.4\%$ over an observation time $480~\omega_p^{-1}$, the same as for our main simulation.

\subsection{Projectile injection}
\label{sec:injection}

The projectiles are introduced individually, one after another. We take two steps to assure that two projectiles are sufficiently separated to avoid cooperative motion among projectiles: after injecting one projectile, we wait for a  time delay of $4.7\omega_t^{-1}$ before injecting the next projectile, and we inject the next projectile from a different site separated by a distance $>8a$.

We now present a simple estimate that demonstrates that a separation $>8a$ provides orders of magnitude of suppression of any cooperative effects. There are two possible mechanisms for interaction among projectiles: direct via pairwise repulsion and indirect via a wake-like disturbance of the target medium. Pairwise repulsion is so small at a distance $>8a$ that it does not even survive our cutoff radius, mentioned above. The wake-like disturbance of the target medium is conveyed by sound waves, the fastest of which is the longitudinal wave. This wave will diminish with distance for two reasons: a $1/r^2$ effect and an exponential decay due to wave damping. The wave damping can be estimated from the sound speed $\approx0.33\omega_ta$, which we determine by analyzing the phonon spectrum for both temperatures, and a damping rate estimated as $\omega_i\geq \nu_t=0.32\omega_t$. Combining these two values, we estimate that a planar longitudinal sound wave is damped by a factor of $1/e$ after a distance of $<1.0a$. Using these values, we can estimate that at a distance of $>8a$, the wake-like disturbances of the medium will diminish by two orders of magnitude due to the $1/r^2$ effect and at least three orders of magnitude due to damping for a total of at least five orders of magnitude. Our use of a launch-site separation of $>8a$ also helps to eliminate any long-lasting ``lane'' effects~\cite{Sutterlin:2009,Sutterlin:2010,Fink:2011,Caliebe:2011,Arp:2011,Schwabe:2011,Zhukhovitskii:2012} that could develop if one projectile were launched from the same site as the previous one.

We do not use periodic boundary conditions because doing so could lead to projectiles wandering too close together. By using a finite simulation box, we can assure that projectiles are always separated by a large multiple of $a$. If instead we used periodic boundary conditions, as a projectile departed on the right side it would be introduced again on the left side, possibly with a separation from the nearest projectile that is $\ll 8a$ due to the cumulative effects of diffusion. We avoid this problem by using finite boundary conditions.

\begin{acknowledgements}
This work was supported by NASA and NSF. We thank S.~D.~Baalrud, W.~D.~S.~Ruhunusiri, and F.~Skiff for helpful discussions.
\end{acknowledgements}


\begin{thebibliography}{20}
\bibitem{note:3}
An examination of  Eqs.~(\ref{EOM}) and (6) shows that using the multiplier is equivalent to adding another Markovian force in addition to that due to the gas, and the resulting motion will have the attributes of thermal motion but at higher temperature.
\bibitem{footnote:2}
These friction constants were calculated using a leading coefficient of 1.26 in the Epstein drag formula~\cite{Liu:03}.
\end{thebibliography}
\end{document}